\documentclass[11pt]{article}
\usepackage{amsmath,amssymb,epsf}

\let\X=\Xi

\def\nn{\nonumber}

\let\bm=\bibitem

\newcommand{\be}{\begin{equation}}
\newcommand{\ee}{\end{equation}}
\def\ba{\begin{array}}
\def\ea{\end{array}}
\def\ft#1#2{{\textstyle{\frac{\scriptstyle #1}{\scriptstyle #2}}}}
\def\fft#1#2{\frac{#1}{#2}}
\def\del{\partial}

\def\sst#1{{\scriptscriptstyle #1}}

\def\td{\tilde}
\def\wtd{\widetilde}

\def\dalemb#1#2{{\vbox{\hrule height .#2pt
        \hbox{\vrule width.#2pt height#1pt \kern#1pt
                \vrule width.#2pt}
        \hrule height.#2pt}}}
\def\square{\mathord{\dalemb{6.8}{7}\hbox{\hskip1pt}}}

\newcommand{\hoch}[1]{$\, ^{#1}$}
\newcommand{\bea}{\begin{eqnarray}}
\newcommand{\eea}{\end{eqnarray}}

\def\0{{\sst{(0)}}}
\def\1{{\sst{(1)}}}
\def\2{{\sst{(2)}}}
\def\3{{\sst{(3)}}}
\def\4{{\sst{(4)}}}
\def\5{{\sst{(5)}}}
\def\6{{\sst{(6)}}}
\def\7{{\sst{(7)}}}
\def\8{{\sst{(8)}}}

\def\R{\rlap{\rm I}\mkern3mu{\rm R}}

\def\R{\rlap{\rm I}\mkern3mu{\rm R}}

\def\R{{{\mathbb R}}}

\def\CP{{{\mathbb C}{\mathbb P}}}

\def\inv#1{\frac{\partial}{\partial #1}}

\def\cP{{\cal P}}

\begin{document}
\begin{flushright}
MIFP-06-04 \\
{\bf hep-th/0602084}\\
February\  2006
\end{flushright}

\begin{center}

{\Large {\bf Separability in Cohomogeneity-2 Kerr-NUT-AdS Metrics}}

\vspace{20pt}

W. Chen, H. L\"u and C.N. Pope

\vspace{20pt}

{\hoch{\dagger}\it George P. \&  Cynthia W. Mitchell Institute
for Fundamental Physics,\\
Texas A\&M University, College Station, TX 77843-4242, USA}

\vspace{40pt}

\underline{ABSTRACT}
\end{center}

    The remarkable and unexpected separability of the Hamilton-Jacobi
and Klein-Gordon equations in the background of a rotating four-dimensional 
black hole played an important r\^ole in the construction of generalisations
of the Kerr metric, and in the uncovering of hidden symmetries associated
with the existence of Killing tensors.  In this paper, we show that
the Hamilton-Jacobi and Klein-Gordon equations are separable in Kerr-AdS
backgrounds in all dimensions, if one specialises the rotation parameters
so that the metrics have cohomogeneity 2.  Furthermore, we show that this
property of separability extends to the NUT generalisations of these
cohomogeneity-2 black holes that we obtained in a recent paper.  In all
these cases, we also construct the associated irreducible rank-2 Killing
tensor whose existence reflects the hidden symmetry that leads to 
the separability.  We also consider some cohomogeneity-1 specialisations
of the new Kerr-NUT-AdS metrics, showing how they relate to previous 
results in the literature.

{\vfill\leftline{}\vfill \vskip 10pt \footnoterule {\footnotesize
Research supported in part by DOE grant
DE-FG03-95ER40917.
}


\newpage
\tableofcontents
\addtocontents{toc}{\protect\setcounter{tocdepth}{2}}

\section{Introduction}

   In order to obtain explicit solutions to the Einstein equations, or
coupled Einstein/matter equations, it is generally necessary to make
simplifying symmetry assumptions about the form the metric.  In some 
cases, where a high degree of symmetry is assumed, this alone can be
sufficient to render the reduced system of equations solvable.  A
typical example is when one considers an ansatz for cohomogeneity-1
metrics, meaning that the remaining metric functions depend non-trivially
on only a single coordinate, and hence the Einstein equations reduce to
a system of ordinary differential equations.
   
   In more complicated circumstances, it may be that symmetries of a
less manifest nature can play an important r\^ole in allowing one to
construct an explicit solution to the Einstein equations.  A nice
example of this kind is provided by the Kerr solution for a
four-dimensional rotating black hole \cite{kerr}.  This is a metric of
cohomogeneity 2, with non-trivial coordinate dependence on both a
radial and an angular variable.  It was observed, after the original
discovery of the solution, that it exhibits the remarkable property,
associated with a ``hidden symmetry,'' of allowing the
separability of the Hamilton-Jacobi equation and the Klein-Gordon
equation.  In fact, it can be shown that the separability is related
to the existence of a 2-index Killing tensor $K_{\mu\nu}$ in
the Kerr geometry, satisfying $\nabla_{(\mu}\, K_{\nu\rho)}=0$.  
By exploiting this property, and conjecturing that it
would continue to hold for the more general situation with a 
cosmological constant, Carter was able to construct the solution
for a four-dimensional asymptotically AdS rotating black hole \cite{carter}.
Further generalisations, with the inclusion, for example, of a NUT parameter,
were subsequently found, again preserving the property of separability.

    It is of considerable interest to investigate the issue of 
separability in other gravitational solutions, including in particular
solutions describing black holes in higher dimensions.  Not only can
this shed light on the existence of hidden symmetries, associated with the
existence of Killing tensors, in the known black hole metrics; it can 
also point the way to constructing more general solutions with additional 
parameters.

    The general $D$-dimensional asymptotically-flat uncharged rotating
black hole was constructed in \cite{myeper}, and the generalisation to
the asymptotically-AdS case with a cosmological constant was
constructed in \cite{gilupapo1,gilupapo2}, extending an earlier result
in \cite{hawhuntay} that was specific to five dimensions.  In both
cases, there are $[(D-1)/2]$ independent rotation parameters $a_i$,
characterising angular momenta in orthogonal spatial 2-planes.  The
general metrics are of cohomogeneity $[D/2]$, with principal orbits
$\R\times U(1)^{[(D-1)/2]}$.  It was shown in \cite{vastpa} that the
Hamilton-Jacobi and Klein-Gordon equations are separable in all odd
dimensions $D$, if one make the specialisation that all $(D-1)/2$
rotation parameters $a_i$ are set equal.  This has the effect of
enhancing the symmetry of the principal orbits from $\R\times
U(1)^{(D-1)/2}$ to $\R\times U((D-1)/2)$, and reducing the cohomogeneity
from $(D-1)/2$ to 1.  In fact, the enhanced manifest symmetry in this
case is already sufficient to permit the separability, without the
need for any additional hidden symmetry.  Indeed, it was shown in
\cite{vastpa} that the Killing tensor in this case is reducible, being a linear
combination of direct products of Killing vectors.  A non-trivial,
irreducible, Killing tensor was found to exist in the case where all
rotation parameters except one are vanishing \cite{chgilupo}.
Irreducible Killing tensors were also shown to exist in the special
case of the five-dimensional asymptotically flat metric of
\cite{myeper}, for arbitrary values of the two rotation parameters
\cite{frosto}.  In the case of rotating AdS black holes in five
dimensions, the separability, and associated irreducible Killing
tensor, were found in \cite{kunluc}.

    A feature common to all the known cases exhibiting the phenomenon
of separability is that the metric in question is of cohomogeneity
$\le 2$.  A natural next step, in the investigation of separability,
is therefore to examine all the $D$-dimensional rotating black holes
under the appropriate specialisation of parameters that reduces their
cohomogeneity from $[D/2]$ to 2. In fact, the $D$-dimensional rotating
AdS black holes with this specialisation were studied recently in 
\cite{chlupo}, and it was shown that they admit a generalisation in which
a NUT parameter is introduced.  Specifically, the specialisation that
reduces the cohomogeneity to 2 is achieved by taking sets of rotation 
parameters to be equal in an appropriate way.  

    In odd dimensions, $D=2n+1$, cohomogeneity 2 is achieved by
dividing the $n=p+q$ rotation parameters $a_i$ into two sets, with 
$p$ of them equal to $a$, and the remaining $q$ parameters equal to $b$.
At the same time, the isometry group enlarges from $\R\times U(1)^{p+q}$ 
to $\R\times U(p)\times U(q)$.  

   In even dimensions $D=2n$, cohomogeneity 2 is achieved by 
instead dividing the $n-1=p+q$ rotation 
parameters into a set of $p$ that are taken to equal $a$, with the remaining
$q$ parameters taken to be zero.  In this case, the isometry group enlarges 
from $\R\times U(1)^{p+q}$ to $\R\times U(p)\times SO(2q+1)$.

   In this paper, we shall show that all these cohomogeneity-2 Kerr-AdS 
metrics have the property that the Hamilton-Jacobi equation and the 
Klein-Gordon equation are separable.  Furthermore, we show that this
property persists when the NUT parameter introduced in \cite{chlupo}
is included.  We also obtain the 2-index Killing tensor $K_{\mu\nu}$ that
is associated with the hidden symmetry responsible for allowing
the equations to separate.  Unlike the case of the further specialisation
to cohomogeneity 1 in odd dimensions that was studied in \cite{vastpa},
in these cohomogeneity-2 cases the Killing tensor is irreducible.

   We also study some further properties of the cohomogeneity-2 Kerr-AdS-NUT
metrics that were obtained in \cite{chlupo}.  In particular, we examine
the case where one adjusts the NUT parameter so that the two adjacent 
roots of the metric function whose vanishing defines the endpoints of 
the range of one of the inhomogeneous coordinates become coincident.  After
appropriate scalings, this limit yields NUT-type metrics of cohomogeneity 1,
which in some special cases coincide with NUT generalisations obtained 
previously in \cite{manste}.

\section{Separability in $D=2n$ Dimensions}\label{evenhjsec}

    It was shown in \cite{chlupo} that if one sets $p$ rotation parameters
$a_i$ equal to $a$, and the remaining $q$ rotation parameters to zero in
general Kerr-AdS metrics of \cite{gilupapo1,gilupapo2} in dimension
$D=2n$, where $p+q=n-1$,
then one can introduce a NUT parameter $L$ to complement the mass parameter
$M$, with the metric being given by 
\bea
ds^2&=&\fft{r^2 +v^2}{X} dr^2 + \fft{r^2 + v^2}{Y} dv^2 -
\fft{X}{r^2 + v^2} \Big(dt - \fft{a^2 - v^2}{a\,\Xi_a}
 (d\psi + A)\Big)^2\label{evenmetric} \\
&& + \fft{Y}{r^2 + v^2}
  \Big(dt - \fft{a^2 + r^2}{a\, \Xi_a} (d\psi + A)\Big)^2
+ \fft{(a^2 + r^2) (a^2 - v^2)}{a^2\Xi_a}d\Sigma_{p-1}^2 +
\fft{r^2 v^2}{a^2} d\Omega_{2q}^2\,.\nn
\eea
Here,  $d\Omega_{2q}^2$ is the metric on the unit sphere $S^{2q}$,
$d\Sigma_{p-1}^2$ is the standard Fubini-Study metric on the ``unit'' 
complex projective space $\CP^{p-1}$ with K\"ahler form $J=\ft12 dA$, 
and the metric functions $X$ and $Y$ are given by 
\bea
X&=& (1 + g^2 r^2) (r^2 + a^2) - \fft{2M\, r}{(r^2 + a^2)^{p-1}\,
r^{2q}}\,,\nn\\
Y&=&(1-g^2 v^2) (a^2 - v^2) - \fft{2L\,v}{(a^2 - v^2)^{p-1}\,
v^{2q}}\,.
\eea
It should be noted that one can replace the unit-sphere metric 
$d\Omega_{2q}^2= \gamma_{ij}\, dx^i dx^j$ 
by any $(2q)$-dimensional Einstein metric normalised to $R_{ij} = (2q-1)\, 
  \gamma_{ij}$, and the Fubini-Study metric $d\Sigma_{p-1}^2 =
   h_{mn}\, dx^m dx^n$ on $\CP^{p-1}$ can be
replaced by any Einstein-K\"ahler $(2p-2)$-metric normalised to
$R_{mn}= 2p h_{mn}$, and one again has a local solution of the Einstein
equations.

   It is not hard to see that the inverse of the metric
(\ref{evenmetric}), which we can write as
$(\del/\del s)^2 \equiv g^{\mu\nu}\, \del_\mu\del_\nu$, is given by
\bea
(r^2 + v^2)\Big(\inv s\Big)^2 &=& X\Big(\inv r\Big)^2 + Y\Big(\inv v\Big)^2 -
\frac{1}{X}\left((r^2 + a^2)\inv t + a \Xi_a \inv \psi
\right)^2\nn\\
&&+\frac{1}{Y}\left((a^2 - v^2)\inv t +
a \Xi_a \inv \psi \right)^2 
+ \; a^2(\frac{1}{r^2} + \frac{1}{v^2})\Big(\inv \Omega\Big)^2
\label{eveninvmet}  \\
&&- \; a^2 \Xi_a \Big(\frac{1}{r^2 + a^2} - \frac{1}{a^2 - v^2}\Big)
h^{m n}
\Big(\partial_m - A_m \inv \psi\Big)\Big(\partial_n - A_n \inv \psi\Big)
\,,\nn
\eea
where $A_m$ are the components of the 1-form $A$, $(\inv \Omega)^2= \gamma^{ij}
\del_i \del_j$ is the 
inverse of the metric on the unit $(2q)$-sphere, and 
$h^{m n}$ are the components of the inverse of the Fubini-Study 
metric on $\CP^{p-1}$.  

\subsection{Separability of the Hamilton-Jacobi equation}

   The covariant Hamiltonian function on the cotangent bundle of the metric
(\ref{evenmetric}) is given by
\be
{\cal H}(\cP_\mu, x^\mu) \equiv \ft12 g^{\mu\nu}\, \cP_\mu\, \cP_\nu\,,
\ee
where $\cP_\mu$ are the canonical momenta conjugate to the coordinates 
$x^\mu$.  In terms of Hamilton's principle function $S$, one has 
$\cP_\mu=\del_\mu S$, and the Hamilton-Jacobi equation is given by
\be
 {\cal H}(\del_\mu S, x^\mu) = -\ft12 \mu^2\,.
\ee
It is evident from (\ref{eveninvmet}) that the Hamilton-Jacobi equation
admits separable solutions of the form
\be
S= -E t + J_\psi\, \psi  + F(r) + G(v) + P + Q\,,\label{evensep}
\ee
where $P$ is a function of the $\CP^{p-1}$ coordinates only, and $Q$ is
a function of the $S^{2p}$ coordinates only. We introduce separation 
constants $K_\Sigma$ and $K_\Omega$ for the functions on these spaces, 
so that
\be
\Big(\fft{\del Q}{\del\Omega}\Big)^2 = K_\Omega^2\,,\qquad
h^{mn}\, (\del_m P - J_\psi\, A_m P)(\del_n P - J_\psi\, A_n P)=
 K_\Sigma^2\,.
\ee

   From the above, we can read off the remaining non-trivial equations for
the functions $F(r)$ and $G(v)$ in (\ref{evensep}), finding
\bea
-2\kappa &=& X {F'}^2 - \fft1{X} \Big(E(r^2+a^2) - a \Xi_a J_\psi\Big)^2 +
  \fft{a^2 K_\Omega^2}{r^2} - \fft{a^2\Xi_a K_\Sigma^2}{r^2+a^2} 
       + \mu^2 r^2\,,\nn\\
2\kappa &=& Y {\dot G}^2 + \fft1{Y} \Big(E(a^2-v^2) - a \Xi_a J_\psi\Big)^2 +
  \fft{a^2 K_\Omega^2}{v^2} + \fft{a^2\Xi_a K_\Sigma^2}{a^2-v^2} 
       + \mu^2 v^2\,,\label{kappaeqn}
\eea
where $F'$ denotes $dF/dr$, $\dot G$ denotes $dG/dv$, and $\kappa$ is the
separation constant associated with the non-trivial hidden symmetry that
permits the separation of the Hamilton-Jacobi equation.

   Note that the separation demonstrated thus far works equally well
if $d\Sigma_{p-1}^2$ is any $(2p-2)$-dimensional Einstein-K\"ahler
metric and $d\Omega_{2q}^2$ is any Einstein metric with the same
scalar curvatures as the $\CP^{p-1}$ and $S^{2q}$ metrics
respectively.  A complete separability, in which the functions $P$ and
$Q$ are themselves fully separated, depends upon the complete
separability of the Hamilton-Jacobi equations in these two spaces.  In
particular, this is possible whenever they are homogeneous spaces, as
is the case for $\CP^{p-1}$ and $S^{2q}$.  Note that in the case of
the Einstein-K\"ahler space, the relevant Hamilton-Jacobi equation is
the one describing a particle of charge $J_\psi$ in geodesic motion, 
with minimal coupling to the potential $A$ whose field
strength is $2J$, where $J$ is the K\"ahler form.

   Following the discussion in \cite{chgilupo}, we note that
associated with the separation constant $\kappa$ is a Poisson
function ${\cal K}$, which Poisson commutes with the Hamiltonian
${\cal H}$.  The function ${\cal K}$ is equal to the separation
constant $\kappa$ if the Hamilton-Jacobi equations are satisfied, and so
we can simply read it off from either of the equations in (\ref{kappaeqn}),
or any linear combination thereof. Thus, for example, from the first
equation in (\ref{kappaeqn}) we may read off
\bea
{\cal K} &=& -\ft12X \cP_r^2 + \fft1{2X} 
  \Big((r^2+a^2)\cP_t + a \Xi_a \cP_\psi\Big)^2 -
  \fft{a^2 \cP_\Omega^2}{2r^2}  +\ft12 r^2g^{\mu\nu}\,\cP_\mu \cP_\nu \nn\\
&&+ \fft{a^2\Xi_a}{2(r^2+a^2)}\, 
   h^{mn}(\cP_m- A_m \cP_\psi)(\cP_n-A_n \cP_\psi)\,,\label{calK0}
\eea
where $\cP_\Omega^2 \equiv \gamma^{ij} \cP_i \cP_j$ and 
$\gamma^{ij}$ is the inverse metric 
on the unit sphere $S^{2q}$.  An alternative way of writing ${\cal K}$,
which puts the $r$ and $v$ coordinates on an equivalent footing, is 
to take the linear combination of the two equations in 
(\ref{kappaeqn}) that eliminates $\mu^2$, yielding
\bea
{\cal K} &=& \fft1{2(r^2+v^2)}\, \Big[ \fft{v^2}{X}\, [( r^2+a^2) \cP_t
   + a\Xi_a \cP_\psi]^2 + \fft{r^2}{Y}\, [(a^2-v^2) \cP_t +
a\Xi_a \cP_\psi]^2 \nn\\
&&\qquad\qquad\qquad
    - v^2 X \cP_r^2 + r^2 Y \cP_v^2\Big]+ \fft{a^2}{2} \Big(\fft1{v^2}
- \fft1{r^2}\Big) \cP_\Omega^2 \nn\\
&&
 + \fft{a^2+r^2-v^2}{2(r^2+a^2)(a^2-v^2)}\, h^{mn} (\cP_m - A_m \cP_\psi)
(\cP_n - A_n \cP_\psi)\,,\label{calK}
\eea

   The function ${\cal K}$ defines a 
St\"ackel-Killing tensor with components $K^{\mu\nu}$, given by
\be
{\cal K} = \ft12 K^{\mu\nu}\, \cP_\mu \cP_\nu\
\ee
Thus the components $K^{\mu\nu}$ can be read off trivially from (\ref{calK0}) 
or (\ref{calK}) by inspection.  The St\"ackel-Killing tensor satisfies
\be
\nabla_{(\mu}\, K_{\nu\rho)}=0\,,
\ee
by virtue of the fact that ${\cal K}$ Poisson commutes with ${\cal H}$.

\subsection{Separability of the Klein-Gordon equation}

    The separability of the Klein-Gordon equation is closely related
to that of the Hamilton-Jacobi equation.  A key observation, which can 
easily be seen from (\ref{evenmetric}), is that
\be
\sqrt{-g} = a^{1-2p-2q}\, \Xi_a^{-p}\, (r^2+a^2)^{p-1}\, (a^2-v^2)^{p-1}\, 
r^{2q}\, v^{2q}\, \sqrt{h}\, \sqrt{\gamma}\, (r^2+v^2)\,.
\ee
Aside from the factor $(r^2+v^2)$, the coordinate dependence of
$\sqrt{-g}$ therefore factorises into a product of a function of $r$, 
a function of $v$, a function of the $S^{2q}$ coordinates and
a function of the $\CP^{p-1}$ coordinates.  Since the Laplacian is given by
\be
\square = \fft1{\sqrt{-g}}\, \del_\mu\Big( \sqrt{-g}\, g^{\mu\nu}\, \del_\nu
           \Big)\,,
\ee
it follows from (\ref{eveninvmet}) that the Klein-Gordon equation 
$\square f = \lambda f$ becomes
\bea
&&\fft1{(r^2+a^2)^{p-1}\, r^{2q}}\, \fft{\del}{\del r}\, 
 \Big( (r^2+a^2)^{p-1}\, 
       r^{2q}\, X\, \fft{\del f}{\del r}\Big)\nn\\
&& +
\fft1{(a^2-v^2)^{p-1}\, v^{2q}}\, \fft{\del}{\del v}\, \Big( (a^2-v^2)^{p-1}\, 
       v^{2q}\, Y\, \fft{\del f}{\del v}\Big) \nn\\
&& - \fft1{X}\Big( (r^2+a^2) \fft{\del}{\del t} +
    a \Xi_a\, \fft{\del}{\del \psi}\Big)^2\, f  +
 \fft1{Y}\Big( (a^2-v^2) \fft{\del}{\del t} +
    a \Xi_a\, \fft{\del}{\del \psi}\Big)^2\, f  \\
&& + a^2\Big(\fft1{r^2} + \fft1{v^2}\Big)\, \fft1{\sqrt{\gamma}}
\, \del_i(\sqrt{\gamma}\, \gamma^{ij}\,\del_j f) \nn\\
&&
-a^2\Xi_a\, 
\Big(\fft1{r^2+a^2} - \fft1{a^2-v^2}\Big)\,\fft1{\sqrt{h}}\, 
   D_m(\sqrt{h} \, h^{mn}\, D_n f) = \lambda\, (r^2+v^2)\, f\,, \nn
\eea
where $D_m\equiv \del_m - A_m\, \del/\del\psi$.  It is manifest that
the equation can be separated by writing $f$ as a product of 
functions of $r$, $v$, $\psi$, the $S^{2q}$ coordinates and the 
$\CP^{p-1}$ coordinates.  Of course the complete separability of
the equation depends upon the fact that one can fully separate the
Klein-Gordon equations on $S^{2q}$ and $\CP^{p-1}$, by virtue of
the homogeneity of these spaces.

\section{Separability in $D=2n+1$ Dimensions}

    It was shown in \cite{chlupo} that if one sets $p$ rotation parameters
$a_i$ equal to $a$, and the remaining $q$ rotation parameters equal to $b$ in
general Kerr-AdS metrics of \cite{gilupapo1,gilupapo2} in dimension
$D=2n+1$, where $p+q=n$,
then one can introduce a NUT parameter $L$ to complement the mass parameter
$M$, with the metric being given by 
\bea
ds^2 &=&\fft{r^2+v^2}{X}\, dr^2 + \fft{r^2+v^2}{Y}\,dv^2 +
  \fft{(r^2+a^2)(a^2-v^2)}{\X_a (a^2-b^2)}\, d\Sigma_{p-1}^2 +
   \fft{(r^2+b^2)(b^2-v^2)}{\Xi_b (b^2-a^2)}\, d\wtd\Sigma_{q-1}^2\nn\\
\!\!&&\!\! +
\fft{a^2b^2}{r^2v^2}\, \Big[ dt - (r^2-v^2)d\phi - r^2 v^2
d\psi -
       \fft{(r^2+a^2)(a^2-v^2)}{a \Xi_a (a^2-b^2)} A -
       \fft{(r^2+b^2)(b^2-v^2)}{b\Xi_b(b^2-a^2)} B\Big]^2\nn\\
&& - \fft{X}{r^2+v^2}\, \Big[ dt + v^2 d\phi -
        \fft{a(a^2-v^2)}{\Xi_a(a^2-b^2)}\, A -
            \fft{b(b^2-v^2)}{\Xi_b (b^2-a^2)}\, B\Big]^2\nn\\
&&+ \fft{Y}{r^2+v^2}\, \Big[ dt - r^2 d\phi -
        \fft{a(r^2+a^2)}{\Xi_a(a^2-b^2)}\, A -
         \fft{b(r^2+b^2)}{\Xi_b (b^2-a^2)}\, B\Big]^2\,,
\label{oddmetric}
\eea
where
\bea
X &=&
  \fft{(1+g^2 r^2)(r^2+a^2)(r^2+b^2)}{r^2} - \fft{2M}{(r^2+a^2)^{p-1}\,
                (r^2+b^2)^{q-1}}\,,\nn\\
Y & =&
   \fft{-(1-g^2 v^2)(a^2-v^2)(b^2-v^2)}{v^2} +
  \fft{2L}{(a^2-v^2)^{p-1}\, (b^2-v^2)^{q-1}}\,.
\eea
Here, $d\Sigma_{p-1}^2$ and $d\wtd\Sigma_{q-1}^2$ are the standard ``unit'' 
metrics on two complex projective spaces $\CP^{p-1}$ and
$\CP^{q-1}$, with K\"ahler forms given locally by $J=\ft12 dA$ and
$\wtd J= \ft12 dB$.  One can also obtain more general solutions by
replacing the complex projective spaces with their
Fubini-Study metrics by any other Einstein-K\"ahler metrics with the
same Ricci scalars.

    One can straightforwardly show that the inverse $(\del/\del s)^2$ 
of the metric (\ref{oddmetric}) is given by
\bea
(r^2 + v^2)\Big(\inv s\Big)^2 
   &=& X \Big(\inv r\Big)^2 + Y \Big(\inv v\Big)^2 +
         \frac{1}{a^2 b^2}\Big(\frac{1}{r^2} + \frac{1}{v^2}\Big)
\Big(\inv \psi\Big)^2
\nonumber \\
&& -\frac{1}{X}\left( r^2 \inv t + \frac{1}{r^2} \inv
         \psi + \inv \phi \right)^2 + \frac{1}{Y}\left( v^2 \inv t +
         \frac{1}{v^2} \inv \psi - \inv \phi \right)^2   \nn\\
&& -(a^2 - b^2) \,\Xi_a \,\Big(\frac{1}{r^2 + a^2} - \frac{1}{a^2 -
         v^2} \Big) \, h^{m n}\, D_m\, D_n \nn\\
&&
 -(b^2 - a^2) \,\Xi_b \,\Big(\frac{1}{r^2 + b^2} - \frac{1}{b^2 -
         v^2}\Big) \, \td h^{k\ell}\, \wtd D_k\, \wtd D_\ell\,,\label{oddinv}
\eea
where
\bea
D_m &\equiv & \partial_m - \frac{a \, A_m}{(a^2-b^2) \Xi_a}\, \Big(\inv \phi -
            a^2 \inv t - \frac{1}{a^2}\inv \psi\Big)\,,\nn\\
\wtd D_k &\equiv & \partial_k - \frac{b \,B_k}{(b^2-a^2) \Xi_b}
         \,  \Big(\inv \phi - b^2 \inv t - \frac{1}{b^2}\inv \psi\Big)\,,
\label{ddef}
\eea
and $h^{mn}$ and $\td h^{k\ell}$ are the inverses of the Fubini-Study
metrics $d\Sigma_{p-1}^2$ and $d\wtd\Sigma_{q-1}^2$ on the complex 
projective spaces $\CP^{p-1}$ and $\CP^{q-1}$.  

\subsection{Separability of the Hamilton-Jacobi equation}

   Following analogous steps to those we described in section \ref{evenhjsec},
it can be seen that the Hamilton-Jacobi equation is separable, if we
write the Hamilton principle function as
\be
S = -E t + J_\psi\, \psi + J_\phi\, \phi + F(r) + G(v) + P + \wtd P\,,
\ee
where $P$ depends only on the coordinates of $\CP^{p-1}$, and $\wtd P$
depends only on the coordinates of $\CP^{q-1}$.  We have separation
constants $K_\Sigma$ and $K_{\wtd\Sigma}$ associated with the two
complex projective space factors.  The Hamilton-Jacobi equations
in these two subspaces themselves describe particles of charges
$q$ and $\td q$ minimally coupled to the vector potentials 
$A$ and $B$ respectively, where
\be
q= \fft{a}{(a^2-b^2)\Xi_a}\, (J_\phi + a^2 E - \fft1{a^2} J_\psi)\,,
\qquad
\td q= \fft{b}{(b^2-a^2)\Xi_b}\, (J_\phi + b^2 E - \fft1{b^2} J_\psi)\,,
\ee
and
\be
h^{mn}\, (\del_m P - q A_m)(\del_n P - q A_n)=K_\Sigma^2\,,\qquad
 \td h^{k\ell}\, (\del_k \wtd P - \td q A_k)
     (\del_\ell \wtd P - \td q A_\ell)=K_{\wtd\Sigma}^2\,.
\ee

  From (\ref{oddinv}), it then follows that there is a further non-trivial
separation constant $\kappa$, leading to the equations
\bea
-2\kappa &=& X {F'}^2 + \fft{J_\psi^2}{a^2 b^2 r^2} - 
       \fft1{X}\, (E r^2 - \fft1{r^2} J_\psi - J_\phi)^2 \nn\\
&&  -\fft{(a^2-b^2) \Xi_a\, K_\Sigma^2}{r^2+a^2} -
   \fft{(b^2-a^2)\Xi_b\, K_{\wtd\Sigma}^2}{r^2+b^2} + \mu^2\, r^2\,,\nn\\
2\kappa &=& Y {\dot G}^2 + \fft{J_\psi^2}{a^2 b^2 v^2} + 
       \fft1{Y}\, (E v^2 - \fft1{v^2} J_\psi + J_\phi)^2 \nn\\
&&  +\fft{(a^2-b^2) \Xi_a\, K_\Sigma^2}{a^2-v^2} +
   \fft{(b^2-a^2)\Xi_b\, K_{\wtd\Sigma}^2}{b^2-v^2} + \mu^2\, v^2\,.
\label{oddkappa}
\eea
We can then read off the associated Poisson function ${\cal K}$ that 
commutes with the Hamiltonian ${\cal H}$, and which takes the constant 
value $\kappa$ upon use of the Hamilton-Jacobi equations.  As in section 
\ref{evenhjsec}, one can organise the expression for ${\cal K}$ in 
different ways, depending on the choice of linear combination of the
two expressions in (\ref{oddkappa}) that one makes.  Thus, for example, 
from the first expression we can write ${\cal K}$ as
\bea
{\cal K} &=& -\ft12 X \cP_r^2 - \fft{1}{2a^2 b^2 r^2}\, \cP_{\psi}^2 + 
       \fft1{2X}\, (r^2 \cP_t + \fft1{r^2} \cP_\psi + \cP_\phi)^2 \nn\\
&& 
  +\fft{(a^2-b^2) \Xi_a}{2(r^2+a^2)}\, \cP_{\Sigma}^2  +
   \fft{(b^2-a^2)\Xi_b}{2(r^2+b^2)}\, \cP_{\wtd\Sigma}^2 +\ft12\, r^2\, 
      g^{\mu\nu}\, \cP_\mu \cP_\nu\,,\label{calKodd0}
\eea
where
\bea
\cP_\Sigma^2 \!\!\!\!&\equiv& \!\!\!\!h^{mn}\, 
         [\cP_m\! -\! \fft{a A_m}{(a^2-b^2)\Xi_a}\, 
 (\cP_\phi -a^2 \cP_t - \fft1{a^2} \cP_\psi)]
[\cP_n\! -\! \fft{a A_n}{(a^2-b^2)\Xi_a}\, 
 (\cP_\phi -a^2 \cP_t - \fft1{a^2} \cP_\psi)]\,, \nn\\
\cP_{\wtd \Sigma}^2 \!\!\!\!&\equiv& \!\!\!\!\td h^{k\ell}\, 
         [\cP_k\! -\! \fft{b B_k}{(b^2-a^2)\Xi_b}\, 
 (\cP_\phi -b^2 \cP_t - \fft1{b^2} \cP_\psi)]
[\cP_\ell\! -\! \fft{b B_\ell}{(b^2-a^2)\Xi_a}\, 
 (\cP_\phi -b^2 \cP_t - \fft1{b^2} \cP_\psi)]\,.
\eea
The components of the associated Killing tensor $K_{\mu\nu}$ can
be read off directly from (\ref{calKodd0}), via ${\cal K}= \ft12 K^{\mu\nu}
 \cP_\mu \cP_\nu$.  Again, as in the even-dimensional case discussed in 
section \ref{evenhjsec}, one can equivalently express ${\cal K}$ in a 
more symmetrical fashion by taking the linear combination of the two
equations in (\ref{oddkappa}) that eliminates $\mu^2$.  

\subsection{Separability of the Klein-Gordon equation}

   As in the case of even dimensions, here too the separability of the 
Klein-Gordon equation is closely related to the separability of the
Hamilton-Jacobi equation.  Again, the key point is that $\sqrt{-g}$ has a
simple form, being proportional to $(r^2+v^2)$ times a product of
functions of $r$, $v$ and the coordinates on the two complex projective
spaces:
\be
\sqrt{-g} = \fft{ab\, r\, v\, \sqrt{h}\, \sqrt{\td h}}{
    |a^2-b^2|^{n-2}\, \Xi_a^{p-1}\, \Xi_b^{q-1}}\,
   (r^2+a^2)^{p-1}\, (r^2+b^2)^{q-1}\, (a^2-v^2)^{p-1}\, 
   (b^2-v^2)^{q-1}\, (r^2+v^2)\,.
\ee
Together with the the expression (\ref{oddinv}) for the inverse metric,
we see that the Klein-Gordon equation $\square f = \lambda f$ assumes 
the manifestly separable form
\bea
&&\fft1{r(r^2+a^2)^{p-1} (r^2+b^2)^{q-1}}\, \inv{r}\, 
      \Big(r(r^2+a^2)^{p-1}(r^2+b^2)^{q-1}\,X\, \fft{\del f}{\del r}\Big)
\nn\\
&&\fft1{v(a^2-v^2)^{p-1} (b^2-v^2)^{q-1}}\, \inv{v}\, 
      \Big(v(a^2-v^2)^{p-1}(b^2-v^2)^{q-1}\,Y\, \fft{\del f}{\del v}\Big)\nn\\
&&-\fft1{X}\, \Big( r^2 \inv t + \fft1{r^2}\, \inv\psi +\inv\phi\Big)^2 f+
\fft1{Y}\, \Big( v^2 \inv t + \fft1{v^2}\, \inv\psi -\inv\phi\Big)^2 f+
\fft1{a^2 b^2}\, \Big(\fft1{r^2} + 
  \fft1{v^2}\Big)\, \fft{\del^2 f}{\del\psi^2}\nn\\
&&-(a^2-b^2)\Xi_a\Big(\fft1{r^2+a^2} - \fft1{a^2-v^2}\Big)\, 
    \fft1{\sqrt h}\, D_m(\sqrt{h}\, h^{mn} D_n f) \nn\\
&& - (b^2-a^2) \Xi_b\Big(\fft1{r^2+b^2} - \fft1{b^2-v^2}\Big)\, 
   \fft1{\sqrt{\td h}}\, \wtd D_k(\sqrt{\td h}\, \td h^{k\ell}\wtd D_\ell f)
  = \lambda (r^2+v^2)f\,. 
\eea
Note that $D_m$ and $\wtd D_k$, defined in (\ref{ddef}), yield gauge
covariant derivatives acting on charged wavefunctions in the two complex
projective spaces, once one separates variables by writing $f$ as a
product of functions of the coordinates.  As in the previous discussions,
the complete separability of the system depends upon the separability
of the Klein-Gordon equations in the complex projective spaces.

\section{Specialisation to NUT Metrics of Cohomogeneity 1}

   The new NUT generalisations of the Kerr-AdS metrics that were found in
\cite{chlupo} all have cohomogeneity 2, and, as we have shown in this 
paper, they all share the feature that the Hamilton-Jacobi equation and
the Klein-Gordon equation are separable in these backgrounds.  It is
also of interest to see how these cohomogeneity-2 Kerr-NUT-AdS metrics 
reduce to certain previously-known solutions under specialisations of the
parameters.  In particular, we shall show that if one applies a limiting
procedure in which the cohomogeneity is reduced from 2 to 1, then the 
resulting metrics include some higher-dimensional NUT metrics that were
obtained in \cite{manste}. As usual, the discussion divides into the
cases of even-dimensional metrics and odd-dimensional metrics.

\subsection{$D=2n$}

   Our starting point is the class of new even-dimensional 
Kerr-NUT-AdS metrics that were obtained in \cite{chlupo}, and which take
the form  given in (\ref{evenmetric}).  The cohomogeneity can be reduced from
2 to 1 by specialising the parameters in such a way that the two adjacent
roots of the function $Y(v)$ that define the range of the $v$ coordinate
become coincident.  Provided the $v$ coordinate is rescaled appropriately 
as the limit is taken, one obtains a non-singular metric that now no longer
has any dependence on the rescaled $v$ coordinate. 

    The function $Y(v)$ acquires a double
root, at $v=v_0$, if the parameters $a$ and $L$ are chosen to satisfy
\bea
L&=&L_0 \equiv \fft{(a^2-v_0^2)^{p+1} v_0^{2q-1}}{
(2q+1) a^2 - (2p+2q+1) v_0^2}\,,\nn\\
g^2 &=& \fft{(2q-1)a^2 - (2p+2q-1) v_0^2}{v_0^2 (
(2q+1) a^2 - (2p+2q +1) v_0^2)}\,.
\eea
In order to approach this limit with an appropriately rescaled $v$ 
coordinate, we define
\be
v=v_0 + \epsilon\, \cos\chi\,,\qquad
L=L_0(1 + \epsilon^2\, c)\,,
\ee
with the constant $c$ given by
\be
c=\fft{a^4 (1-4q^2) + 2a^2(2q-1)(2p+2q+1)v_0^2 +
(1-4(p+q)^2) v_0^4}{2(a^2-v_0^2)^2 v_0^2}\,,
\ee
where $\epsilon$ will shortly be sent to zero. The function $Y$ under 
this limit becomes
\be
Y = \epsilon^2 Y_0 \sin^2\chi\,,
\ee
where
\be
Y_0=\fft{2 (a-v_0)^2c}{(2p+2q+1) v_0^2-(2q+1)a^2}\,.
\ee
In order for the metric (\ref{evenmetric}) to be nonsingular in the
limit, we must also make the coordinate transformations
\be
\psi\rightarrow \fft{a\,\Xi_a}{\epsilon\, Y_0} \td\psi\,,\qquad
t\rightarrow t + \fft{a^2 - v_0^2}{\epsilon\, Y_0}\, \psi\,.
\ee
Sending $\epsilon$ to zero, the metric (\ref{evenmetric}) then becomes
\bea
ds^2 &=& - \fft{X}{r^2 + v_0^2} (dt + \fft{2v_0}{Y_0} \cos\chi\, d\psi
-\fft{a^2-v_0^2}{a\,\Xi_a} A)^2 + \fft{r^2 + v_0^2}{X} dr^2\nn\\
&&+\fft{(r^2+v_0^2)}{Y_0}(d\chi^2 + \sin^2\chi\, d\psi^2) +
\fft{r^2 + a^2}{a^2\, \Xi_a} d\Sigma_{p-1}^2 + \fft{r^2 v_0^2}{a^2}
d\Omega_{2q}^2\,.\label{evennut}
\eea

   The metrics (\ref{evennut}) are contained within a rather general class
of cohomogeneity-1 NUT metrics that were obtained in \cite{manste}. The
case $p=1$ and $q=0$ reduces to the standard Taub-NUT-AdS metric
in four dimensions.

\subsection{$D=2n+1$}

    In odd dimensions, our starting point is the cohomogeneity-2 
Kerr-NUT-AdS metrics found in \cite{chlupo}, and presented in 
equation (\ref{oddmetric}).

    Proceeding in an analogous fashion to the discussion we gave in 
even dimensions, we first consider the conditions under 
which $Y$ has a double root, at $v=v_0$.  This happens when
the constants $L$, $a$ and $b$ are chosen such that
\bea
L&=&L_0\equiv \fft{(a^2-v_0^2)^p (b^2-v_0^2)^q (1- g^2 v_0^2)}{
2v_0^2}\,,\nn\\
g^2&=& \fft{a^2b^2 + (a^2(q-1) + b^2(p-1))v_0^2 -(p+q-1) v_0^4}{
(a^2 q + b^2 p -(p+q) v_0^2)v_0^4}\,.
\eea
Next, we deform away slightly from the double root, and introduce a new
coordinate $\chi$ in place of $v$:
\be
v=v_0 + \epsilon\, \cos\chi\,,\qquad
L=L_0(1 + \epsilon^2\, c)\,,
\ee
where the constant $c$ is given by
\bea
c&=&\fft{2}{(a^2-v_0^2)^2 (b^2 - v_0^2)^2}
\Big(-2a^2 b^2 (a^2q + b^2 p) -
(a^4q(q-1) + b^4p(p-1)\\
&&\qquad+ 2a^2b^2 (pq - 2p-2q)) v_0^2
+2(p+q)(a^2(q-1) + b^2(p-1)) v_0^4\nn\\
&&\qquad\qquad - (p+q)(p+q-1) v_0^6\Big)\,.\nn
\eea
The function $Y$ in this limit becomes
\be
Y=\epsilon^2 Y_0\, \sin^2\chi\,,
\ee
where
\be
Y_0=\fft{(a^2-v_0^2)^2(b^2-v_0^2)^2\,c}{v_0^4((p+q)
v_0^2 - a^2 q - b^2 p)}\,.
\ee
Making the further coordinate transformation
\be
t\rightarrow t - \fft{v_0^2\phi}{\epsilon\,Y_0}\,,\qquad
\phi\rightarrow \fft{\phi}{\epsilon\,Y_0}\,,\qquad
\psi\rightarrow -\fft{\phi}{v_0^2\epsilon\, Y_0} -
\fft{\psi}{v_0^4}\,,
\ee
we can now obtain a smooth limit in which $\epsilon$ is sent to
zero, for which the metric (\ref{oddmetric}) becomes
\bea
ds^2&=&-\fft{X}{r^2 + v_0^2}\Big[dt + \fft{2v_0}{Y_0}\cos\chi\,
d\phi -\fft{a(a^2-v_0^2)}{\Xi_a(a^2 - b^2)}A -
-\fft{b(b^2-v_0^2)}{\Xi_b(b^2 - a^2)}B\Big]^2\nn\\
&&+\fft{a^2b^2}{r^2v_0^2} \Big[ dt +
\fft{2v_0}{Y_0}\cos\chi\, d\phi +
\fft{r^2}{v_0^2}(d\psi + \fft{2v_0}{Y_0}\cos\chi\,d\phi)
-\fft{(r^2+a^2)(a^2-v_0^2)}{a\Xi_a (a^2-b^2)} A\nn\\
&&-\fft{(r^2 +b^2)(b^2-v_0^2)}{b\Xi_b (b^2-a^2)} B\Big]^2
+\fft{r^2+ v_0^2}{X} dr^2 + \fft{r^2 + v_0^2}{Y_0}(d\chi^2 +
\sin^2\chi\, d\phi^2)\nn\\
&&+\fft{(r^2+a^2)(a^2-v_0^2)}{\Xi_a(a^2-b^2)} d\Sigma_{p-1}^2
+\fft{(r^2+b^2)(b^2-v_0^2)}{\Xi_b(b^2-a^2)} d\wtd\Sigma_{q-1}^2
\,.
\eea
This metric is contained within the class of cohomogeneity-1 NUT
generalisations that were considered in \cite{manste}.

\section{Conclusions}

    The separability of the Hamilton-Jacobi
and Klein-Gordon equations in the background of a rotating four-dimensional 
black hole played an important r\^ole in the construction of generalisations
of the Kerr metric, and in the uncovering of hidden symmetries associated
with the existence of Killing tensors.  In this paper, we have shown that
the Hamilton-Jacobi and Klein-Gordon equations are separable in Kerr-AdS
backgrounds in all dimensions, if one specialises the rotation parameters
so that the metrics have cohomogeneity 2.  Furthermore, we have shown that this
property of separability extends to the NUT generalisations of these
cohomogeneity-2 black holes that we obtained in \cite{chlupo}.  In all
these cases, we also constructed the associated irreducible rank-2 Killing
tensor whose existence reflects the hidden symmetry that leads to 
the separability.  We also considered some cohomogeneity-1 specialisations
of the new Kerr-NUT-AdS metrics, and showed how they relate to previous 
results in the literature \cite{manste}.

   The results on separability that we have obtained in this paper
raise the interesting question of whether it might extend to the 
higher-dimensional rotating black holes with more general choices for
the rotation parameters, and thus having cohomogeneity larger than 2.
This question is currently under investigation.

\end{document}